# Spatial entanglement of fermions in one-dimensional quantum dots

**Ivan P. Christov** [1,2]

1  Physics Department, Sofia University, 1164 Sofia, Bulgaria;

[2]  Institute of Electronics, Bulgarian Academy of Sciences, 1784 Sofia, Bulgaria

**Abstract:** The time dependent quantum Monte Carlo method for fermions is introduced and applied for calculation of entanglement of electrons in one-dimensional quantum dots with several spin-polarized and spin-compensated electron configurations. The rich statistics of wave functions provided by the method allows one to build reduced density matrices for each electron and to quantify the spatial entanglement using measures such as quantum entropy by treating the electrons as identical or distinguishable particles. Our results indicate that the spatial entanglement in parallel-spin configurations is rather small and it is determined mostly by the quantum nonlocality introduced by the ground state. By contrast, in the spin-compensated case the outermost opposite-spin electrons interact like bosons which prevails their entanglement, while the inner shell electrons remain largely at their Hartree-Fock geometry. Our findings are in a close correspondence with the numerically exact results, wherever such comparison is possible.

## 1. Introduction

During the past few decades there has been an increasing interest in developing new models and computational tools to address the fundamental and practical challenges related to quantum correlations and entanglement in connection with their potential applications in the newly emerging quantum technologies [1]. The properties of composite systems of quantum particles are expected to play an important role in information processing as well as in devises for manipulating systems of atoms and molecules. While various algebraic operator methods have been used to characterize entanglement in spin systems [2-4] an efficient approach to assess the spatial entanglement in many-body quantum systems together with its temporal evolution is still lacking. It is well known that the correlated non-relativistic particle motion described by the time-dependent Schrödinger equation (SE) is tractable for only a limited number of cases. While solvable numerically for few particles in 1D and 2D the direct numerical solution of SE scales exponentially with system size and is therefore beyond the capabilities of the today computers. That exponential-time scaling is usually attributed to the non-local quantum effects which result from the dependence of the wave function $\Psi(\mathbf{r}_1,...,\mathbf{r}_N,t)$ on the coordinates of all interacting particles. The standard approaches to ameliorate the workload are to reduce the many-body SE to a set of coupled one-body equations, where the most prominent are the mean-field approaches: Hartree-Fock (HF) method [5] and the Density Functional theory (DFT) [6]. That reduction however occurs at the price of neglecting the detailed fluctuating forces between the electrons and replacing those by averages thus totally ignoring the dynamic quantum correlations in HF while DFT reduces the many-body problem to a single-body problem of non-interacting electrons moving in an effective exchange-correlation potential which is generally unknown and suffers self-interaction issues. More accurate but more computationally expensive are the multi-configuration time dependent Hartree–Fock [7], and the full configuration-interaction [8] methods. Another approach which has gained much attention lately is the density matrix renormalization group method [9] which has allowed one to treat with good accuracy correlated 1D many-body problems however its application to higher dimensions and arbitrary potentials has been challenging so far.

A different class of methods to tackle the quantum many body problem includes the quantum Monte Carlo methods [10] which allowed to accurately calculate the electronic structure of atoms, molecules, nano-structures, and condensed-matter systems at a fully correlated level. For example the diffusion quantum Monte Carlo (DMC) uses random particles (walkers) whose evolution towards the ground state of the system involves a combination of diffusion and branching, which however prevent its use for real-time dependent processes where the causality is of primary importance. Also the artificial nature of the many-body wave function in configuration space used in DMC prevents the calculation of some important quantities other than the energy. The recently introduced time-dependent quantum Monte Carlo (TDQMC) method [11-13] employs concurrent ensembles of walkers and wave functions for each electron where each wave function is associated with a separate walker (particle-wave dichotomy) and these evolve in physical space-time where no initial guess for the many-body wave-function is needed. Recently we have applied the TDQMC method to analyze the ground state preparation for simple bosonic systems with several electrons in one and two dimensions with a good tradeoff between scaling and accuracy [14]. In this work we apply the TDQMC method to several interacting fermions in one dimensional quantum dots where the obtained reduced density matrices allow us to quantify the entanglement between the electrons considered as identical or distinguishable particles. We consider some proof-of-principle aspects of the TDQMC method for fermions rather than practical matters concerning quantum dots. Entanglement in two-electron quantum dots, atoms and molecules has been studies elsewhere [15-18].

## 2. Methods

The TDQMC method transforms the standard Hartree-Fock (HF) equations into a set of coupled stochastic equations capable of describing the correlated particle motion. That transformation is based on the physical assumption that the modulus-square of the one-body wave function in coordinate (physical) space may be considered as an envelope (or kernel density estimation) of the distribution of a finite number of particles (walkers). In this way for each electron in an atom a large set of one-body wave functions which reside in physical space-time is created where each wave function responds to the multi-core potential due to both the nucleus and the walkers of the rest of the electrons. The crucial point in this picture is that it allows for each walker for a given electron to interact with the walkers of any other electron through weighted Coulomb potential thus naturally incorporating the quantum non-locality. Then, from the evolution of the walker distributions one can evaluate quantum observables without resorting to the many-body wave function. Formally, we start from the HF equation for the i-th from a total of N electrons, within a one-determinant ansatz [19]:

$$i\hbar \frac{\partial}{\partial t}\varphi_i(\mathbf{r}_i,t) = \left[ -\frac{\hbar^2}{2m}\nabla_i^2 + V_{en}(\mathbf{r}_i) + V_{ee}^{HF}(\mathbf{r}_i,t) \right]\varphi_i(\mathbf{r}_i,t) , \tag{1}$$

where $V_{en}(\mathbf{r}_i)$ is the electron-nuclear potential, and the HF electron-electron potential reads:

$$V_{ee}^{HF}(\mathbf{r}_i,t) = V_{ee}^{H}(\mathbf{r}_i,t) + V_{ee}^{X}(\mathbf{r}_i,t), \tag{2}$$

where:

$$V_{ee}^{H}(\mathbf{r}_i,t) = \sum_{j \neq i}^{N} \int d\mathbf{r}_j V_{ee}(\mathbf{r}_i - \mathbf{r}_j) |\varphi_j(\mathbf{r}_j,t)|^2 \qquad (3)$$

is the Hartree potential, and $V_{ee}^{X}(\mathbf{r}_i,t)$ is the exchange potential:

$$V_{ee}^{X}(\mathbf{r}_i,t) = -\sum_{j \neq i}^{N} \delta_{s_i,s_j} \int d\mathbf{r}_j V_{ee}(\mathbf{r}_i - \mathbf{r}_j) \varphi_i(\mathbf{r}_j,t) \varphi_j^{*}(\mathbf{r}_j,t) \varphi_j(\mathbf{r}_i,t) / \varphi_i(\mathbf{r}_i,t), \qquad (4)$$

where $\varphi_i(\mathbf{r},t)$ satisfy the orthonormality property $\int \varphi_i(\mathbf{r},t)\varphi_j^{*}(\mathbf{r},t)d\mathbf{r} = \delta_{i,j}$, and the indices $s_i, s_j$ denote the spins of the corresponding electrons. The inequality $j \neq i$ in the sums of Equations (3) and (4) stresses the fact that even though the self-interaction between the electrons is naturally canceled in the HF approximation it is not present also in the Hartree approximation where there is no exchange potential [19]. It is known that the wave functions $\varphi_i(\mathbf{r},t)$ of Equation (1) variationally minimize the system energy:

$$E^{HF} = \sum_{i=1}^{N} \left[ -\frac{\hbar^2}{2m} \int \varphi_i^{*}(\mathbf{r}_i,\tau) \nabla_i^2 \varphi_i(\mathbf{r}_i,\tau) d\mathbf{r}_i + \int V_{en}(\mathbf{r}_i) |\varphi_i(\mathbf{r}_i,\tau)|^2 d\mathbf{r}_i \right] + E_{ee}^{H} + E_{ee}^{X}, \qquad (5)$$

where the Hartree and the exchange energies read:

$$E_{ee}^{H} = 0.5 \sum_{i \neq j}^{N} \iint V_{ee}(\mathbf{r}_i - \mathbf{r}_j) |\varphi_i(\mathbf{r}_i,\tau)|^2 |\varphi_j(\mathbf{r}_j,\tau)|^2 d\mathbf{r}_i d\mathbf{r}_j, \qquad (6)$$

and

$$E_{ee}^{X} = -0.5 \sum_{i \neq j}^{N} \delta_{s_i,s_j} \iint V_{ee}(\mathbf{r}_i - \mathbf{r}_j) \varphi_i(\mathbf{r}_j,\tau) \varphi_j^{*}(\mathbf{r}_j,\tau) \varphi_j(\mathbf{r}_i,\tau) \varphi_i^{*}(\mathbf{r}_i,\tau) d\mathbf{r}_i d\mathbf{r}_j, \qquad (7)$$

respectively.

It is known that the Hartree-Fock approximation does not account for the dynamic electron-electron correlations beyond those due to the exchange interaction. In order to correct for this in the TDQMC methodology we replace the HF wave function for each electron $\varphi_i(\mathbf{r},t)$ by a family of slightly different wave functions $\varphi_i(\mathbf{r},t) \to \varphi_i^k(\mathbf{r},t)$; k=1,…,M [11-13] which allows to further lower the system energy below the HF level. This is accomplished by applying a stochastic windowing to the distribution $|\varphi_j(\mathbf{r}_j,t)|^2$ in the Hartree potential $V_{ee}^H(\mathbf{r}_i,t)$ of Equation (3) by using a "window" function $K[\mathbf{r}_j,\mathbf{r}_j^k(t),\sigma_{j,i}]$ centered at certain trajectory $\mathbf{r}_j^k(t)$ which samples the distribution given by $|\varphi_j(\mathbf{r}_j,t)|^2$. The parameters $\sigma_{j,i}$ determine the widths of those "windows" such that the product $|\varphi_j(\mathbf{r}_j,t)|^2 K[\mathbf{r}_j,\mathbf{r}_j^k(t),\sigma_{j,i}]$ is different for each separate trajectory $\mathbf{r}_j^k(t)$. In this way, for each electron, Equations (1)-(4) are transformed into a set of M Schrödinger -like equations for the different replicas $\varphi_i^k(\mathbf{r}_i,t)$ of the initial HF wave function $\varphi_i(\mathbf{r}_i,t)$, each one attached to separate trajectory $\mathbf{r}_j^k(t)$ (particle-wave dichotomy, [13]):

$$i\hbar \frac{\partial}{\partial t}\varphi_i^k(\mathbf{r}_i,t) = \left[ -\frac{\hbar^2}{2m}\nabla_i^2 + V_{en}(\mathbf{r}_i) + V_{eff}^k(\mathbf{r}_i,t) \right. \tag{8}$$

$$\left. -\sum_{j\neq i}^N \delta_{s_i,s_j} \int d\mathbf{r}_j V_{ee}(\mathbf{r}_i-\mathbf{r}_j)\varphi_i^k(\mathbf{r}_j,t)\varphi_j^{k*}(\mathbf{r}_j,t)\varphi_j^k(\mathbf{r}_i,t)/\varphi_i^k(\mathbf{r}_i,t) \right] \varphi_i^k(\mathbf{r}_i,t)$$

where $\int \varphi_i^k(\mathbf{r},t)\varphi_j^{k*}(\mathbf{r},t)d\mathbf{r} = \delta_{i,j}$; i=1,…,N, k=1,…,M, and where:

$$V_{eff}^k(\mathbf{r}_i,t) = \sum_{j\neq i}^N \frac{1}{Z_{j,i}^k} \sum_l^M V_{ee}\left[\mathbf{r}_i,\mathbf{r}_j^l(t)\right] K\left[\mathbf{r}_j^l(t),\mathbf{r}_j^k(t),\sigma_{j,i}\right] \tag{9}$$

is the effective electron-electron interaction potential represented as a Monte-Carlo (MC) convolution which incorporates the spatial quantum nonlocality by allowing each walker for a given electron to interact with a group of walkers of any other electron. In fact, it is seen from Equation (9) that the effective potential "seen" by the k-th wave for the i-th electron involves the interactions with a number of walkers which belong to the j-th electron which lie within the non-local length $\sigma_{j,i}$ around $\mathbf{r}_j(t)$ [12-14]. For the Gaussian kernel we have:

$$K\left[\mathbf{r}_j, \mathbf{r}_j^k(t), \sigma_{j,i}\right] = \exp\left(-\frac{\left|\mathbf{r}_j - \mathbf{r}_j^k(t)\right|^2}{2\sigma_{j,i}^2}\right), \tag{10}$$

which determines the weighting factor in Eq.(9) to be:

$$Z_{j,i}^k = \sum_{l=1}^{M} K\left[\mathbf{r}_j^l(t), \mathbf{r}_j^k(t), \sigma_{j,i}\right] \tag{11}$$

As seen from Equations (10), (11) the limit $\sigma_{j,i} \to \infty$ where $K\left[\mathbf{r}_j, \mathbf{r}_j^k(t), \sigma_{j,i}\right] \to 1$ recovers the mean field approximation as opposed to the local interaction where $\sigma_{j,i} \to 0$ and $K\left[\mathbf{r}_j, \mathbf{r}_j^k(t), \sigma_{j,i}\right] \to \delta\left(\mathbf{r}_j - \mathbf{r}_j^k(t)\right)$. It is clear therefore that $\sigma_{j,i}$ may serve as variational parameters to minimize the system energy between these two limiting cases.

The connection between the trajectories $\mathbf{r}_i^k(t)$ and the waves $\varphi_i^k(\mathbf{r}_i, t)$ is given by the walker's velocities (de Broglie-Bohm equation. e.g. in [20]):

$$\mathbf{v}_i^k(t) = \frac{\hbar}{m} \mathrm{Im}\left[\frac{\nabla_i \varphi_i^k(\mathbf{r}_i, t)}{\varphi_i^k(\mathbf{r}_i, t)}\right]_{\mathbf{r}_i = \mathbf{r}_i^k(t)} \tag{12}$$

for real-time propagation, and:

$$d\mathbf{r}_i^k(\tau) = \mathbf{v}_i^{Dk} d\tau + \mathbf{\eta}_i(\tau)\sqrt{\frac{\hbar}{m} d\tau} \tag{13}$$

for imaginary-time propagation, where the drift velocity is:

$$\mathbf{v}_i^{Dk}(\tau) = \frac{\hbar}{m}\left[\frac{\nabla_i \varphi_i^k(\mathbf{r}_i, \tau)}{\varphi_i^k(\mathbf{r}_i, \tau)}\right]_{\mathbf{r}_i = \mathbf{r}_i^k(\tau)}, \tag{14}$$

and $\boldsymbol{\eta}(\tau)$ is Markovian stochastic process (see also the appendix in Ref.[21]). The striking similarity between the drift velocity of Equation (14) and the de Broglie-Bohm Equation (12) comes from the fact that both equations describe drift-diffusion processes in imaginary and in real time, respectively. It is seen that although the individual waves guide the corresponding walkers through Equation (12), the TDQMC method solves coupled one-body Hartree-Fock-like equations (8) instead of using quantum potentials as done in Bohmian mechanics [20].

Following the particle-wave dichotomy described above the system energy can be calculated conveniently using both particle trajectories and wave functions:

$$E = \frac{1}{M}\sum_{k=1}^{M}\left[\sum_{i=1}^{N}\left[-\frac{\hbar^2}{2m}\frac{\nabla_i^2 \varphi_i^k(\mathbf{r}_i^k)}{\varphi_i^k(\mathbf{r}_i^k)} + V_{en}(\mathbf{r}_i^k)\right] + \sum_{i>j}^{N}V_{ee}(\mathbf{r}_i^k - \mathbf{r}_j^k)\right]_{\substack{\mathbf{r}_i^k = \mathbf{r}_i^k(\tau) \\ \mathbf{r}_j^k = \mathbf{r}_j^k(\tau)}} \quad (15)$$

$$-\frac{0.5}{M}\sum_{k=1}^{M}\sum_{i\neq j}^{N}\delta_{s_i,s_j}\iint d\mathbf{r}_i d\mathbf{r}_j V_{ee}(\mathbf{r}_i - \mathbf{r}_j)\varphi_i^k(\mathbf{r}_j,t)\varphi_j^{k*}(\mathbf{r}_j,t)\varphi_j^k(\mathbf{r}_i,t)\varphi_i^{k*}(\mathbf{r}_i,t)$$

During the preparation of the ground state of the quantum system the initial Monte Carlo ensembles of walkers and guide waves propagate in imaginary time ($\tau$) toward steady state, in accordance with Equations (8)-(14), for different values of the nonlocality parameters $\sigma_{j,i}$ until the energy of Equation (15) displays a minimum. Another alternative to account for the exchange effects is to apply a short-range screening to the interaction potential in order to modify the repulsion between the same-spin electrons [22].

Considering the ensemble of waves $\varphi_i^k(\mathbf{r}_i,t)$ delivered by the TDQMC as random variables one can build a reduced density matrix for the i-th electron which may serve as the variance-covariance matrix in Hilbert space which carries important statistical information [13], [23]:

$$\rho_i(\mathbf{r}_i,\mathbf{r}_i',t) = \frac{1}{M}\sum_{k=1}^{M}\varphi_i^{k*}(\mathbf{r}_i,t)\varphi_i^k(\mathbf{r}_i',t) \quad (16)$$

For example the density matrix of Equation (16) allows one to easily calculate the spatial entanglement of a given electron state which may serve also as a good measure for the overall accuracy of the calculation (see e.g. [24]). Without entering the ongoing debate on entanglement witnesses, for opposite-spin electrons where there are no exchange terms in HF and in TDQMC equations, we employ the linear quantum entropy for distinguishable (non-identical) particles as a conventional measure for the spatial entanglement [25]:

$$S^i_{L\uparrow\downarrow}(t) = 1 - Tr(\rho_i^2) = 1 - \int \rho_i^2(\mathbf{r}_i, \mathbf{r}_i, t) d\mathbf{r}_i , \quad (17)$$

while for N same-spin electrons the component to the entanglement which reflects the trivial minimum correlation that is due to the anti-symmetrization of the wave function can be eliminated [26-28], yielding:

$$S^i_{L\uparrow\uparrow}(t) = 1 - NTr(\rho_i^2) = 1 - N\int \rho_i^2(\mathbf{r}_i, \mathbf{r}_i, t) d\mathbf{r}_i \quad (18)$$

This definition ensures that for the wave functions used in HF approximation (or in general for any Slater rank 1 many-body state [29-30]) the linear entropy should vanish.

For indistinguishable (identical) particles, the spatial part of the 2N-body wave function can be represented in the simplest case as a product of normalized spin-up and spin-down Slater determinants [10]:

$$\Psi_i(\mathbf{r}_1, \mathbf{r}_2, ..., \mathbf{r}_{2N}, t) = D_i^{\uparrow}(\mathbf{r}_1, \mathbf{r}_2, ..., \mathbf{r}_N, t) D_i^{\downarrow}(\mathbf{r}_{N+1}, \mathbf{r}_{N+2}, ..., \mathbf{r}_{2N}, t) . \quad (19)$$

Thus, the entanglement between e.g. spin-up only electrons can be estimated using the reduced density matrix, averaged over the configurations provided by the TDQMC algorithm:

$$\rho_i^{\uparrow}(\mathbf{r}, \mathbf{r}', t) = \frac{1}{M}\sum_k^M \int D_i^{k\uparrow}(\mathbf{r}, \mathbf{r}_2, ..., \mathbf{r}_N, t) D_i^{k\uparrow *}(\mathbf{r}', \mathbf{r}_2, ..., \mathbf{r}_N, t) d\mathbf{r}_2 ... d\mathbf{r}_N , \quad (20)$$

where $D_i^{k\uparrow}$ are spin-up Slater determinants composed by the individual wave functions.

### 3. Results

As an example here we calculate the ground state of a quantum dot with parabolic core potential $V_{en}(\mathbf{r}_i) = \omega^2 r_i^2 / 2$ and with soft-core electron-electron Coulomb repulsion [31]:

$$V_{ee}\left[\mathbf{r}_i, \mathbf{r}_j\right] = \frac{e^2}{\sqrt{r^2 + a^2}}, \tag{21}$$

where $r \equiv |\mathbf{r}_i - \mathbf{r}_j|$.

Within the formalism of Section 2 the degree of spatial correlation and hence the spatial entanglement is controlled in TDQMC by the quantum nonlocal length $\sigma_{j,i}$ where, for bound electrons, higher $\sigma_{j,i}$ lead to lower correlation (entanglement) between the i-th and the j-th electron, and vice-versa. Since the spatial extend of the electron cloud for the j-th electron is determined by the standard deviation $s_j$ of the corresponding MC ensemble, the nonlocal length $\sigma_{j,i}$ is expected to be close to $s_j$:

$$\sigma_{j,i} = \alpha_{j,i} \cdot s_j; \qquad j,i=1,\ldots,N, \tag{22}$$

where $\alpha_{j,i}$ may now serve as the variational parameters to minimizing the energy.

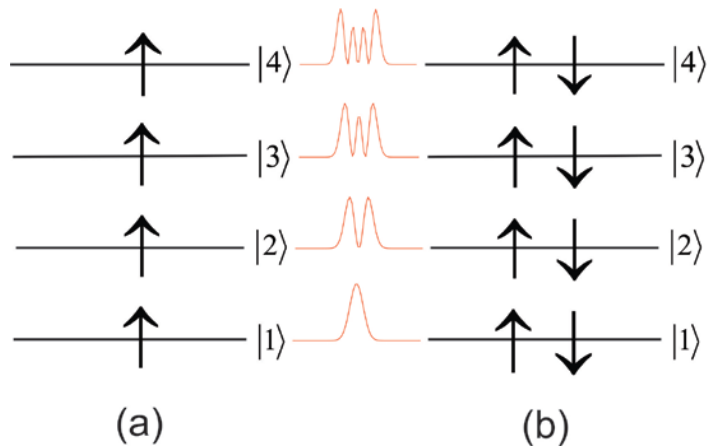

**Figure 1.** Energy level diagrams for spin-polarized -(**a**) and spin-compensated -(**b**) electrons in 1D quantum dot. The moduli-square of the corresponding spatial orbitals are drawn with red.

Since for parallel-spin electrons the eigenstates are orthogonal to each other their overlap is small and hence the dynamic correlation between such states is expected to be smaller as compared to the correlation between opposite-spin electrons. Here we consider consider 1D quantum dots in two basic configurations: the one is a spin-polarized configuration where each energy level is filled with just one same-spin electron as seen in Figure 1(a), and spin-compensated configuration where each level is occupied by two opposite spin electrons (Figure 1(b)).

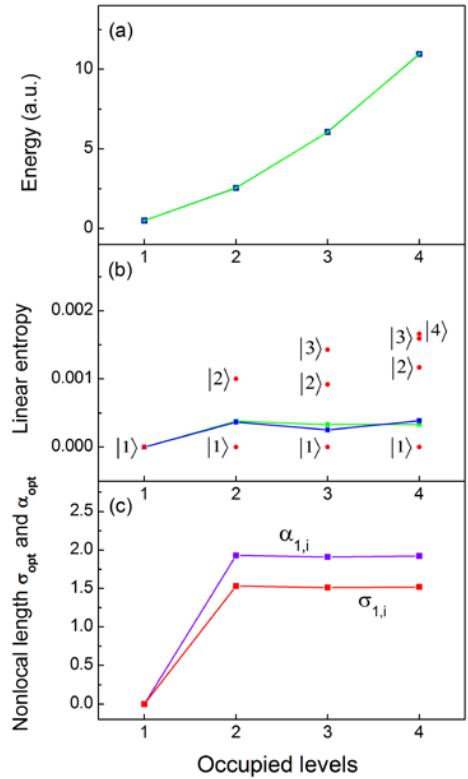

**Figure 2.** Energy - (**a**), linear entropy - (**b**), and nonlocality parameters $\sigma_{1,i}$ ($\alpha_{1,i}$) for the ground state $|1\rangle$ - (**c**), for electron configurations with up to four parallel-spin electrons (Figure 1(a)). Blue lines –TDQMC results, green lines –numerically exact results, red dots in (**b**) – linear entropy for distinguishable electrons.

We start with the ground state and the three excited states of a total of four same-spin electron configurations (Figure 1(a)) where due to the orthogonality of the spatial wave functions the dynamic correlations between the equal-spin electrons are expected to be rather small. Starting from preliminary calculated Hartree-Fock wave functions, after 200 steps of imaginary-time propagation of Equations (8)-(11) and Equations (13), (14), for $\omega=1$, and for different values of the variational parameter $\alpha_{j,i}$ of Eq.(22), we find the energy minima for 1, 2, 3, and 4 occupied levels in succession. Figure 2(a) shows with green line these energies which are in a very good correspondence with the numerically exact energies (blue line) obtained from the direct numerical solution of the Schrödinger equation for up to four electrons in one spatial dimension. Our calculations reveal that accuracy of 3 significant digits for the energy can be attained by varying $\alpha_{1,i}$, while $\alpha_{2,i}$, $\alpha_{3,i}$, and $\alpha_{4,i}$ are set to infinity which practically keeps the ground state (level 1) at its Hartree-Fock geometry. The optimal values of $\alpha_{1,i}$ for the ground level $|1\rangle$ are shown in Figure 2c for 2, 3, and 4 electrons also showing that both $\alpha_{1,i}$ and the nonlocal length $\sigma_{1,i}$ are almost independent on the number of electrons, except for one electron at the ground state where there is no e-e interaction and $\alpha_{1,i}$ is set to zero. The degree of entanglement for the four configurations of Figure 1(a) is quantified by the linear quantum entropy of Equation (18) as shown in Figure 2(b). There are two distinct cases: the first one where the electrons are considered identical (blue line) to be compared with the result from the exact numerical solution of the Schrödinger equation (green line). It is seen that the linear entropy in this case remains almost constant in close agreement with the exact numerical result. The second case plotted with red dots in Figure 2(b) depicts the linear entropy for the different electrons considered as distinguishable particles with the density matrix of Equation (17). It is seen that the linear entropy for the distinguishable electrons increases due to the screening effect of the inner electrons which causes larger fluctuations in the shape of the outer wave functions

For the filled-shell configuration of Figure 1(b) each level contains two opposite-spin electrons whose wave functions overlap in space almost completely and therefore these interact more like bosons [14]. It is therefore reasonable to calculate only the spatial entanglement that is due to same-shell states which is expected to exceed significantly the entanglement of the same-spin electrons at different levels. For the configuration of Figure 1(b) we have found that the major source of entanglement is the Coulomb repulsion between the two outermost electrons. For the ground state (level 1) where only two opposite-spin electrons are present in the vicinity of the core the system energy exhibits well-defined minimum as function of the nonlocal length $\sigma_{1,1}^{\uparrow}=\sigma_{1,1}^{\downarrow}$ as seen in Figure 3(a). When adding electrons at the higher levels the corresponding wave functions acquire zeros which is a source of larger fluctuations of the energy as seen from Figure 3 (b)-(d) where the red curves represent the polynomial least-squares fit for better visualization of the energy minimum.

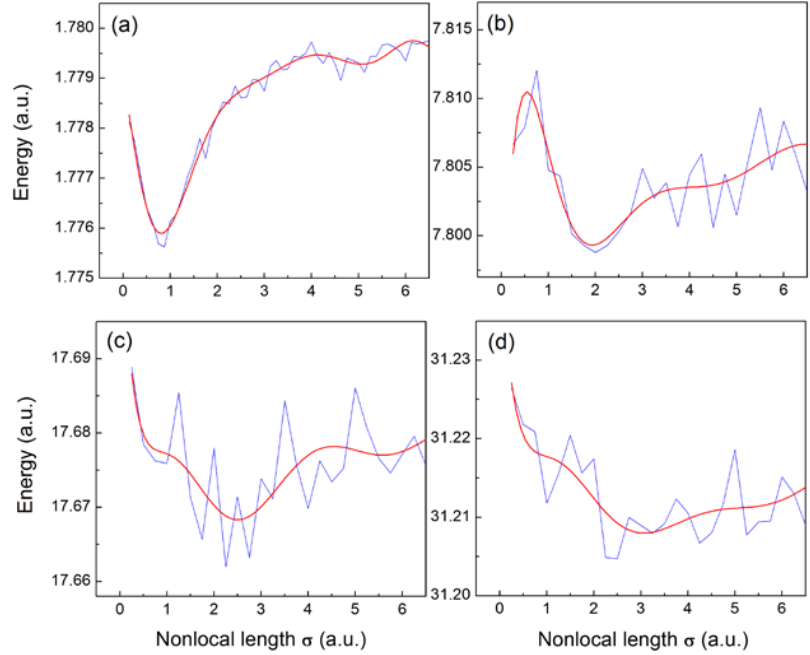

**Figure 3.** Energy of N-electron 1D quantum dot as function of the nonlocal length $\sigma_{N,N}$, for N=2 -(**a**), N=4 -(**b**), N=6 -(**c**), and N=8 -(**d**).

Note that in the process of adding correlated electrons at the outer-shells the inner-shell electrons remain largely intact due to their stronger localization (confinement) to the core. Therefore, to a good approximation, the inner-shell wave functions need not be recalculated and these may remain at their self-consistent Hartree-Fock configurations. Figure 4(a) depicts the system energy (blue line) which almost perfectly matches the numerically exact energies (green line) obtained using the standard DMC method [10]. The linear entropy predicted by the TDQMC method decreases when adding new excited states to the electron configuration (blue line in Figure 4(b)) which contrasts the case of bosonic quantum dots where it increases for more electrons at the ground state [14]. This behavior is confirmed also by the numerically exact results (green line) for levels 1 and 2 (two and four identical electrons) and can be explained by the orthogonality of the wave functions in the fermionic calculation which causes weaker interaction for the outer-shell electrons.

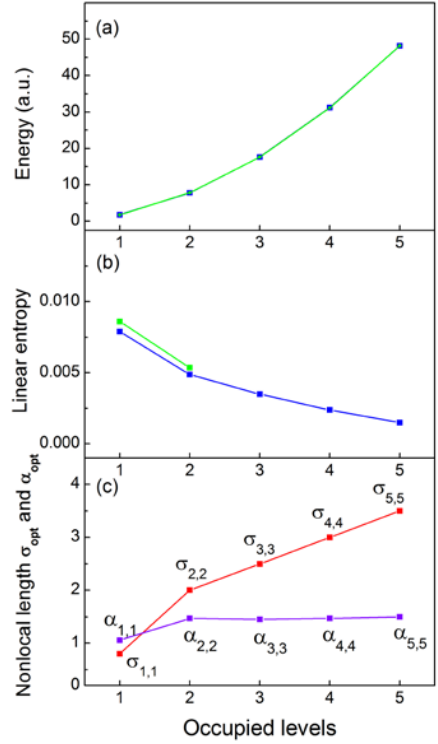

**Figure 4.** Energy - (**a**), linear quantum entropy - (**b**), and nonlocality parameters $\sigma_{1,i}$ ($\alpha_{1,i}$) for the outermost level $|i\rangle = |1\rangle,...,|5\rangle$ - (**c**), for electron configurations with up to five filled shells (Figure 1(b)). Blue lines –TDQMC results, green lines –numerically exact results.

## 4. Conclusions

In conclusion, we have calculated the correlated ground state and excited states of 1D quantum dots with spin-polarized and spin-compensated electron configurations with up to five energy levels (ten electrons) within the time-dependent quantum Monte Carlo framework. By minimizing variationally the system energy with respect to the quantum nonlocality the optimal set of wave functions that describes each electron is found, which allows further to calculate the reduced density matrices for the different electrons and hence to quantify the entanglement they exhibit due to their mutual interactions. Using the linear quantum entropy as a measure for the entanglement it was found that for a fully spin-polarized electron configuration the stochastic windowing applied to the ground state alone is sufficient to recover the entanglement of the excited states in good agreement with the exact numerical result. For the spin-compensated electron system the entanglement that is due to the interaction of the two outermost opposite-spin electrons is dominant while the inner shells remain largely at their Hartree-Fock states. An essential advantage of the method is that it allows one to conceive quantum particles as identical as well as distinguishable objects. The theory presented here may find useful applications in treating quantum correlation effects in composite quantum systems such as molecules, clusters, and solid-state materials.

**Funding:** This research is based upon work supported by the Air Force Office of Scientific Research under award number FA9550-19-1-7003, and by the Bulgarian Ministry of Education and Science as a part of National Roadmap for Research Infrastructure, grant number D01-401/18.12.2020 (ELI ERIC BG).